\documentclass[journal=cgdefu,manuscript=article]{achemso}
\usepackage[version=3]{mhchem} 
\usepackage{multirow}
\usepackage{arydshln}
\author{Daniel A. Mayoh}
\email{d.mayoh.1@warwick.ac.uk}
\author{George D. A. Wood}
\author{Samuel Holt}
\author{Grady Beckett}
\author{Emily J. L. Dekker}
\author{Martin R. Lees}
\author{Geetha Balakrishnan}
\affiliation[Warwick University]
{Department of Physics, University of Warwick, Coventry, UK}
\email{g.balakrishnan@warwick.ac.uk}

\title[]{Effects of Fe deficiency and Co substitution in polycrystalline and single crystals of Fe$_{3}$GeTe$_{2}$}

\abbreviations{}

\begin{document}

\begin{abstract}

\noindent Fe$_{3}$GeTe$_{2}$ is a two-dimensional van der Waals material with a ferromagnetic ground state and a maximum transition temperature $T_{\mathrm{c}}\sim225$~K. However, when Fe$_{3}$GeTe$_{2}$ is synthesized lower values of {$T_{\mathrm{c}}$} are often reported. This is attributed to a deficiency in the Fe at the 2c site in the crystal structure. Here we investigate the effect of Fe deficiency and the substitution of Co for Fe on the magnetic properties of this system. We have synthesized both polycrystalline material and single crystals by chemical vapor transport and the flux method, with the largest crystals obtained using the flux method. Cobalt substitution at the Fe site is found to significantly reduce the magnetic transition temperature. Crystals of Fe$_{3}$GeTe$_{2}$ grown by chemical vapor transport with $\sim 8\%$ excess Fe in the starting materials display an optimum Fe content and magnetic transition temperature.

\end{abstract}


\section{Introduction}

Van der Waals (vdW) bonded magnetic materials are currently shaping the field of two-dimensional (2D) materials science~\cite{Mak:2019}. The presence of van der Waals bonding in these materials allows them to be cleaved down to monolayers and combined with other 2D materials to create novel heterostructures without concern for lattice matching. By modifying these heterostructures the electrical, optical and magnetic properties can be fine tuned and in some cases unique magnetic phases can be realized~\cite{Burch:2018}. Many magnetic vdW materials have been discovered to host an assortment of magnetic orders such as ferromagnetism (e.g. \ce{CrB3}~\cite{Zhang:2019}, \ce{CrI3}~\cite{Li:2020,Liu:2020}, \ce{CrGeTe3}~\cite{Carteaux:1995,Han:2019} and \ce{CrSiTe3}~\cite{Carteaux:1991, Cao:2020}), antiferromagnetism (e.g. MnBi$_{2}$Te$_{4}$~\cite{Deng:2020}, MnP(S/Se)$_{3}$~\cite{Susner:2017} and CrCl$_{3}$~\cite{Kuhlow:1982}), helimagnetism (e.g. NiBr$_{2}$~\cite{Tokunaga:2011} and NiI$_{2}$~\cite{Kurumaji:2013}) and frustrated magnetism (e.g. RuCl$_{3}$~\cite{Kim:2015}).

Fe$_{3}$GeTe$_{2}$ is a 2D magnetic material which has been generating significant attention as it was the first vdW material found to be both magnetic and metallic. It also offers a significant step up in Curie temperature ($T_{\mathrm{c}}\sim225$~K) from previous magnetic vdW materials.~\cite{Deiseroth:2006,Chen:2013,Li:2018} Much research has been directed towards tuning the magnetic properties of this material by layer-thickness dependence studies \cite{fei:2018}, varying the \ce{Fe} content \cite{may:2016}, \ce{Co} substitution~\cite{tian:2019,Hwang:2019}, and the investigation of the anomalous Hall effect~\cite{Kim:2018}, planar topological Hall effect~\cite{you:2019}, Kondo lattice physics~\cite{Zhang:2018}, anisotropy magnetostriction effect~\cite{Zhuang:2016} and ionic liquid voltage gating~\cite{deng:2018}. In bulk Fe$_{3}$GeTe$_{2}$ there have been some conflicting reports over whether the ground state of Fe$_{3}$GeTe$_{2}$ should be antiferromagnetic or ferromagnetic~\cite{yi:2016,kong:2020,jang:2020}. 

In its stoichiometric form, there should be inter-layer antiferromagnetism in Fe$_{3}$GeTe$_{2}$, however, \ce{Fe} vacancies in this system, which correspond to more hole charge carriers, produce a ferromagnetic ground state~\cite{jang:2020}. This observation agrees with the majority of the literature, in which samples are frequently reported to be \ce{Fe} deficient and ferromagnetic~\cite{may:2016}. It has also been suggested that a antiferromagnetic to ferromagnetic transition in Fe$_{3}$GeTe$_{2}$ can be induced by electron doping~\cite{jang:2020}.

\begin{figure}[t]
\centering
\includegraphics[width=\textwidth]{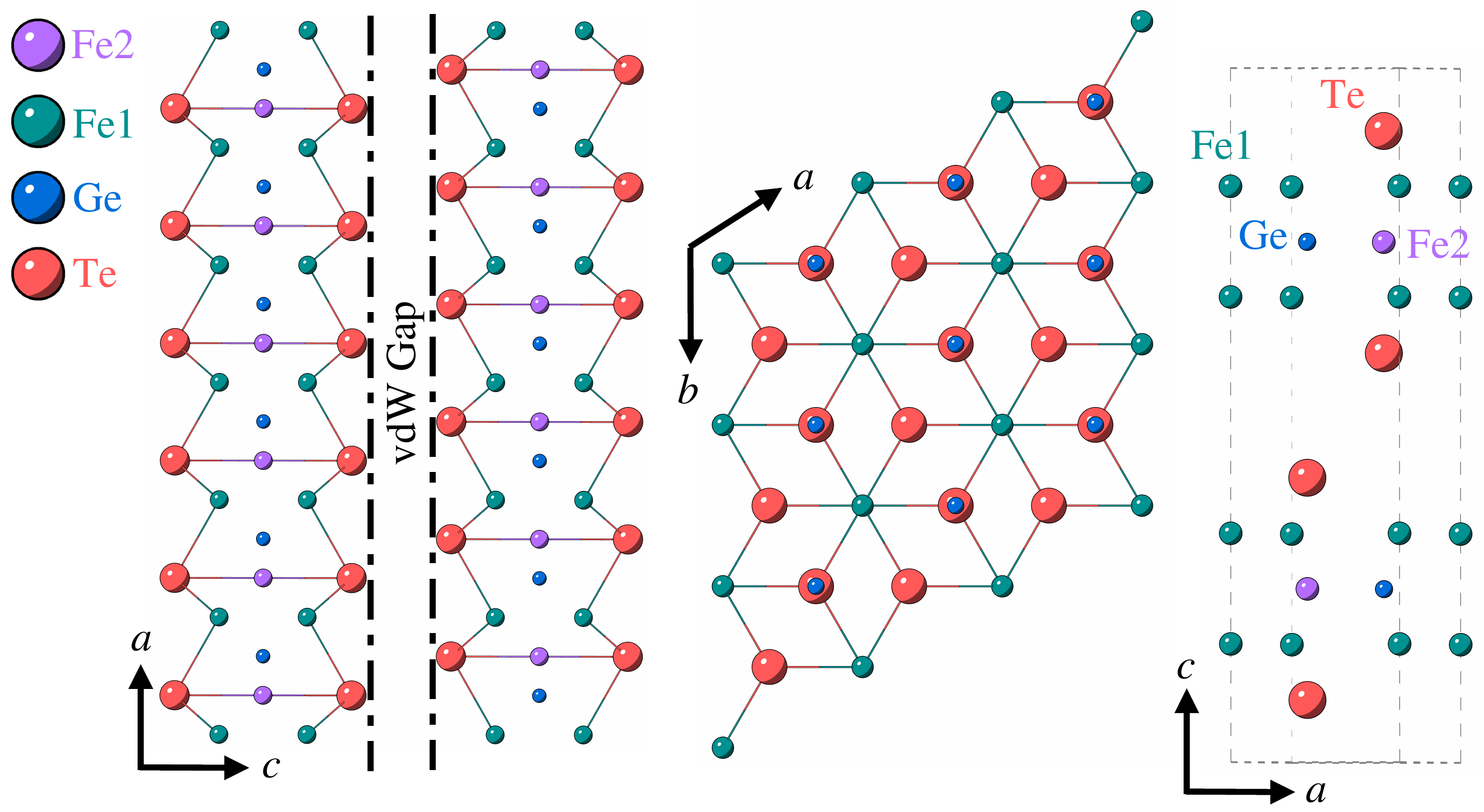}
\caption{Crystal structure of  Fe$_{3}$GeTe$_{2}$ along several crystallographic orientations. The vdW bonding is most easily observed when looking along the $ac$ plane as indicated by the dashed-dotted lines between the Te layers. The hexagonal nature of the structure is clearly observable when looking perpendicular to the $ab$ plane. The partial occupancy of Fe at the 2c (Fe2) site is found to be instrumental in determining the nature of the magnetic transition in  Fe$_{3}$GeTe$_{2}$. The Fe atoms found on the 4c (Fe1) and 2c (Fe2) sites are shown in green and purple, respectively. The Te and Ge atoms are indicated in red and blue, respectively. The dashed lines indicate the unit cell of Fe$_{3}$GeTe$_{2}$.}
\label{FIG: Crystal Structure}
\end{figure}

In addition to the highly tunable nature of the magnetic properties of Fe$_{3}$GeTe$_{2}$, there have been several studies concentrating on skyrmions and the spin textures which have been observed in this material. Magnetic bubble-like formations were first observed in Fe$_{3}$GeTe$_{2}$ when a magnetic \ce{Ni} tip of a scanning tunneling microscope (STM) was held close to a freshly cleaved surface of Fe$_{3}$GeTe$_{2}$~\cite{nguyen:2018}. Lorentz transmission electron microscopy (TEM) measurements further revealed magnetic bubble-like textures in Fe$_{3}$GeTe$_{2}$ which were claimed to be skyrmions~\cite{Ding:2020}. These bubbles could be tuned in size with applied magnetic field. The presence of a topological Hall effect has recently been observed in Fe$_{3}$GeTe$_{2}$, this signal is frequently take as the hallmark of skyrmions~\cite{you:2019,chowdhury:2020}. Since Fe$_{3}$GeTe$_{2}$ crystallizes in the centrosymmetric hexagonal space group $P6_{3}/mmc$ (No. 194) as shown in Figure~\ref{FIG: Crystal Structure}, the mechanism for the stabilization of skyrmions in bulk Fe$_{3}$GeTe$_{2}$ is currently unclear. Skyrmion-like magnetic bubbles have only been observed in Fe$_{3}$GeTe$_{2}$ in a thin film regime or at the interface of the material, as is the case for the STM study, where it is conceivable that inversion symmetry can be broken leading to a Dzyaloshinskii-Moriya (DM) interaction~\cite{Dzyaloshinky:1958}. The observation of Néel type skyrmions at an oxide interface in Fe$_{3}$GeTe$_{2}$ gives further confirmation that the DM interaction is allowing the stabilization of skyrmions in this material~\cite{park:2021}. Should skyrmions be realized in bulk Fe$_{3}$GeTe$_{2}$, the interactions stabilizing them cannot include the DM interaction due to the centrosymmetric nature of the crystal structure.

We have undertaken a detailed investigation into the dependence of the magnetic properties on the Fe content in the Fe$_{3}$GeTe$_{2}$ materials. We have synthesized both polycrystalline and single crystals of Fe$_{3}$GeTe$_{2}$ and Co substituted materials, Fe$_{3-y}$Co$_{y}$GeTe$_{2}$ for $y = 0.3,~0.6,~0.9,~1.2~\mathrm{and}~1.5$. Investigations of the structure and properties of the polycrystalline and single crystals produced have been undertaken using x-ray diffraction and magnetic measurement techniques. Composition analysis, including estimates of the Fe content in these materials, reveal a direct correlation with the observed magnetic transitions. We also find that an excess of Fe is necessary in the synthesis of Fe$_{3}$GeTe$_{2}$ to optimize the  Fe content and therefore achieve the highest $T_{\mathrm{c}}$.


\section{Experimental details}

Polycrystalline materials were synthesized by solid state reaction. Single crystal growths were carried out by either chemical vapor transport (CVT) or the flux method using Te flux. All the sample preparation techniques are discussed in more detail in the following section. To determine the phase purity and crystal structure of the synthesized polycrystalline materials, powder x-ray diffraction was performed at room temperature using a Panalytical Empyrean diffractometer (Bragg-Brentano geometry) with a Cu K$_{\alpha 1}$ and K$_{\alpha2}$ source and a solid state PIXcel detector. Rietveld refinements were carried out on the observed diffraction patterns using the TOPAS academic v6.0 software suite~\cite{Coelho:jo5037}. The quality of the single crystals obtained was investigated by Laue x-ray imaging using a Photonic Science Laue camera. 

The chemical composition was determined using a ZEISS GeminiSEM 500 which was used to perform energy dispersive x-ray spectroscopy (EDX). The magnetic properties of both the single crystals and the polycrystalline materials were measured using a Quantum Design Magnetic Property Measurement System (MPMS), superconducting quantum interference device (SQUID) magnetometer. Measurements were made in the temperature range 2-300~K in various applied magnetic fields in the zero-field-cooled (ZFC) and field-cooled (FC) modes. 

\section{Material Synthesis}\label{SEC: Synthesis}

\subsection{Polycrystalline synthesis}

Polycrystalline materials of Fe$_{3}$GeTe$_{2}$ and Fe$_{3-y}$Co$_{y}$GeTe$_{2}~(y=0.3,~0.6,~0.9,~1.2,~1.5)$ were synthesized by the solid state reaction. Stoichiometric quantities of high purity elements in powder form, Fe (STREM Chemicals, Inc., 99.99\%), Co(Alfa Aesar, 99.998\%), Ge (Acros Organics, 99.999\%), and Te (Alfa Aesar, 99.99\%) were ground together inside an argon filled glove box and transferred into quartz ampoules. The evacuated quartz ampoules were held at $675~^{\circ}$C for 10 days with heating/cooling rates of $\sim60-100~^{\circ}$C/h. The resulting grey/black powders were ground and investigated for phase formation. 

\subsection{Single crystal growth}

Two different techniques were employed to obtain single crystals of the various compositions listed, chemical vapor transport, and the flux method using excess Te.

\begin{figure}[t]
\centering
\includegraphics[width=\textwidth]{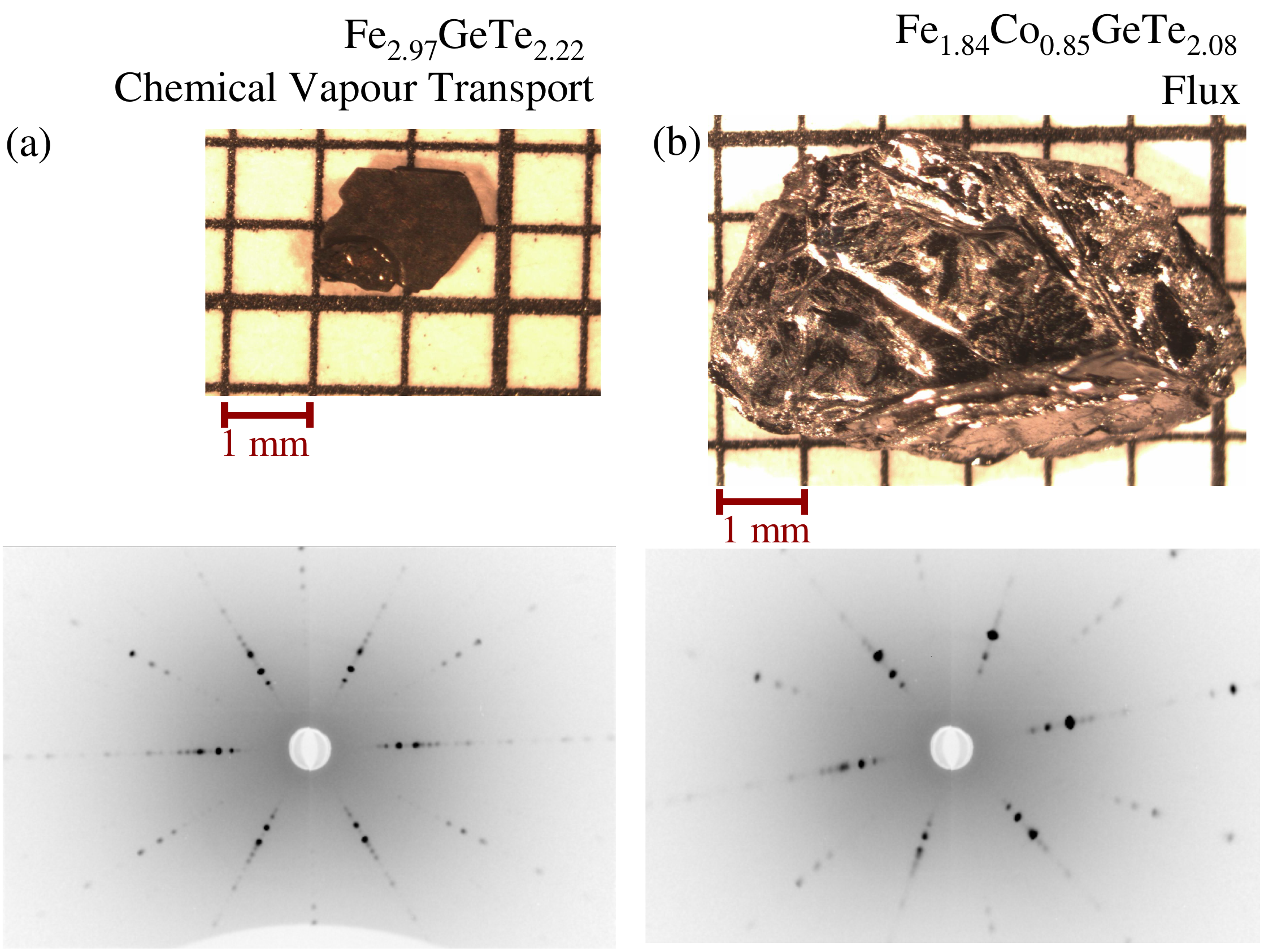}
\caption{A typical as-grown crystal of (a) Fe$_{3.25}$GeTe$_{2}$ prepared using CVT and (b) Fe$_{3-y}$Co$_{y}$GeTe$_{2}$ grown by the flux method. The compositions of the single crystals are those estimated by EDX analysis. Below each picture is a corresponding Laue diffractogram displaying the 6-fold symmetry expected observing along the $c$~axis of the crystal.}
\label{FIG: Crystal Pictures}
\label{FIG: Laue}
\end{figure}

For the CVT process, growths using two different transport agents, I$_{2}$ and TeCl$_{4}$, were attempted. Quantities of Fe, Ge, Co and Te powders (see Table~\ref{Tab: EDX}) along with 5 mg/cm$^{3}$ of the transport agent were sealed in quartz ampoules. The growth of crystals was carried out by holding the source and the sink ends of each tube at different temperatures in a two zone furnace for two weeks, before cooling to room temperature. Several different temperature profiles were used, 750-675~$^{\circ}$C for the hot end and 700-650~$^{\circ}$C for the cold end. In some of the CVT growths, in addition to the platelets which have been identified as corresponding to the desired Fe$_{3}$GeTe$_{2}$ crystals, other needle-like and pyramidal-shaped structures corresponding to other phases of Fe-Ge-Te were also obtained. A photograph of a typical crystal platelet obtained from a CVT growth is shown in Figure~\ref{FIG: Crystal Pictures}(a).  

For the crystal growth by the flux method, using excess Te (powder) as the flux, mixtures of varying nominal compositions, (see Table~\ref{Tab: EDX}), were used. The mixtures were placed in an alumina crucible which was then sealed in a quartz ampoule under vacuum. To enable the capture of the crystals during a subsequent centrifuging process, a small amount of quartz wool was placed over the alumina crucible inside the ampoule. The tubes were heated to 1000~$^{\circ}$C and cooled at the rate of 3~$^{\circ}$C/h to 675~$^{\circ}$C, at which temperature the tubes were removed from the furnace and centrifuged to remove the excess Te flux~\cite{Ding:2020,may:2016}. An example of a crystal grown by the flux method is shown in Figure~\ref{FIG: Crystal Pictures}(b).

In general the thickness and size of the crystals grown by the flux method were found to be larger than those obtained by CVT. The average size of the CVT grown crystals is $2 \times 2$~mm$^2$ while those obtained by the flux method are typically $7 \times 5$~mm$^2$.

\section{Results and discussion}

\subsection{Laue diffraction}

X-ray back reflection Laue patterns were taken on the crystals to check for crystalline quality. Typical Laue photographs of isolated platelets of Fe$_{3}$GeTe$_{2}$ and Fe$_{3-y}$Co$_{y}$GeTe$_{2}$, mounted with the $ab$ plane perpendicular to the x-ray beam are shown in Figures~\ref{FIG: Laue}(a) and \ref{FIG: Laue}(b). These display the 6-fold symmetry expected observing along the $c$~axis of the crystals. 

\subsection{Powder x-ray diffraction}

\begin{figure}[t]
\centering
\includegraphics[width=\textwidth]{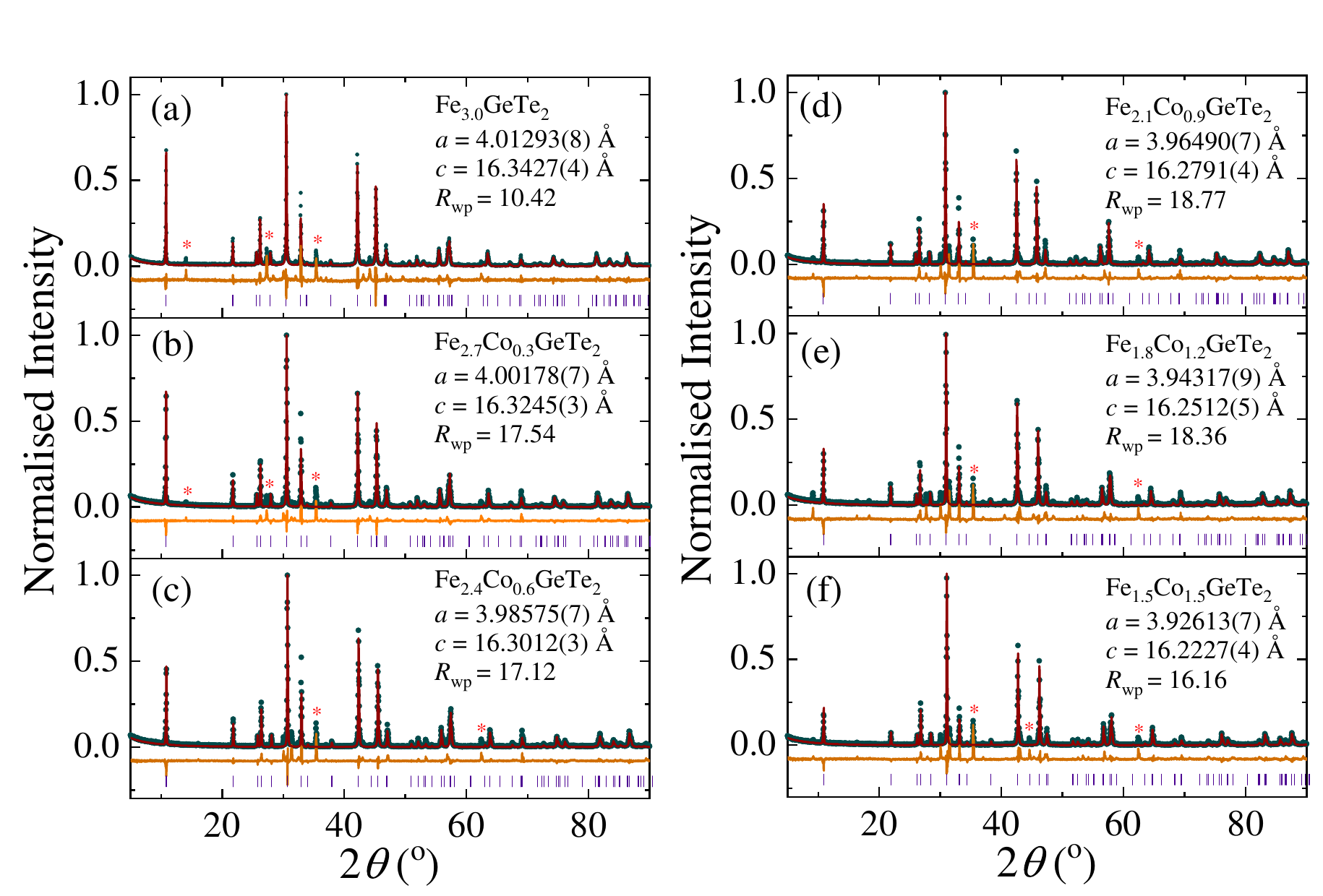}
\caption{Room temperature powder x-ray diffraction patterns of polycrystalline samples with the nominal compositions (a) Fe$_{3.0}$GeTe$_{2}$, (b) Fe$_{2.7}$Co$_{0.3}$GeTe$_{2}$, (c) Fe$_{2.4}$Co$_{0.6}$GeTe$_{2}$, (d) Fe$_{2.1}$Co$_{0.9}$GeTe$_{2}$, (e) Fe$_{1.8}$Co$_{1.2}$GeTe$_{2}$, and (f) Fe$_{1.5}$Co$_{1.5}$GeTe$_{2}$. Shown are the experimental profiles (green circles) and the Rietveld refinement (red solid line) made using the hexagonal $P6_{3}/mmc$ space group with the calculated difference (orange solid line). The purple bars indicate the expected positions of the Bragg peaks. The physical and crystallographic parameters obtained for Fe$_{3.0}$GeTe$_{2}$ are given in Table.~\ref{Tab: pxrd FGT}. Impurity peaks are denoted by a red asterisk.}
\label{FIG: PXRD}
\end{figure}

Phase purity and structural analysis were carried out using powder x-ray diffraction on the polycrystalline materials synthesized. The diffraction patterns obtained could be indexed to the hexagonal space group $P6_{3}/mmc$ (No.~194). Figure~\ref{FIG: PXRD} shows the observed diffraction patterns for the various nominal starting compositions of the Fe-Ge-Te and Fe-Co-Ge-Te powders and the Rietveld refinements of the observed patterns obtained using the TOPAS software suite. The typical physical and crystallographic parameters obtained for one of the compositions synthesized, Fe$_{3}$GeTe$_{2}$, are given in Table~\ref{Tab: pxrd FGT} along with the $R_{\mathrm{wp}}$ values indicating the quality of the fits. The lattice parameters obtained from the fits to the powder diffraction patterns are also shown in Figure~\ref{FIG: PXRD}. The lattice parameters obtained are largely in agreement with those reported for these materials.~\cite{Deiseroth:2006} 

\begin{figure}[t]
\centering
\includegraphics[width=1\textwidth]{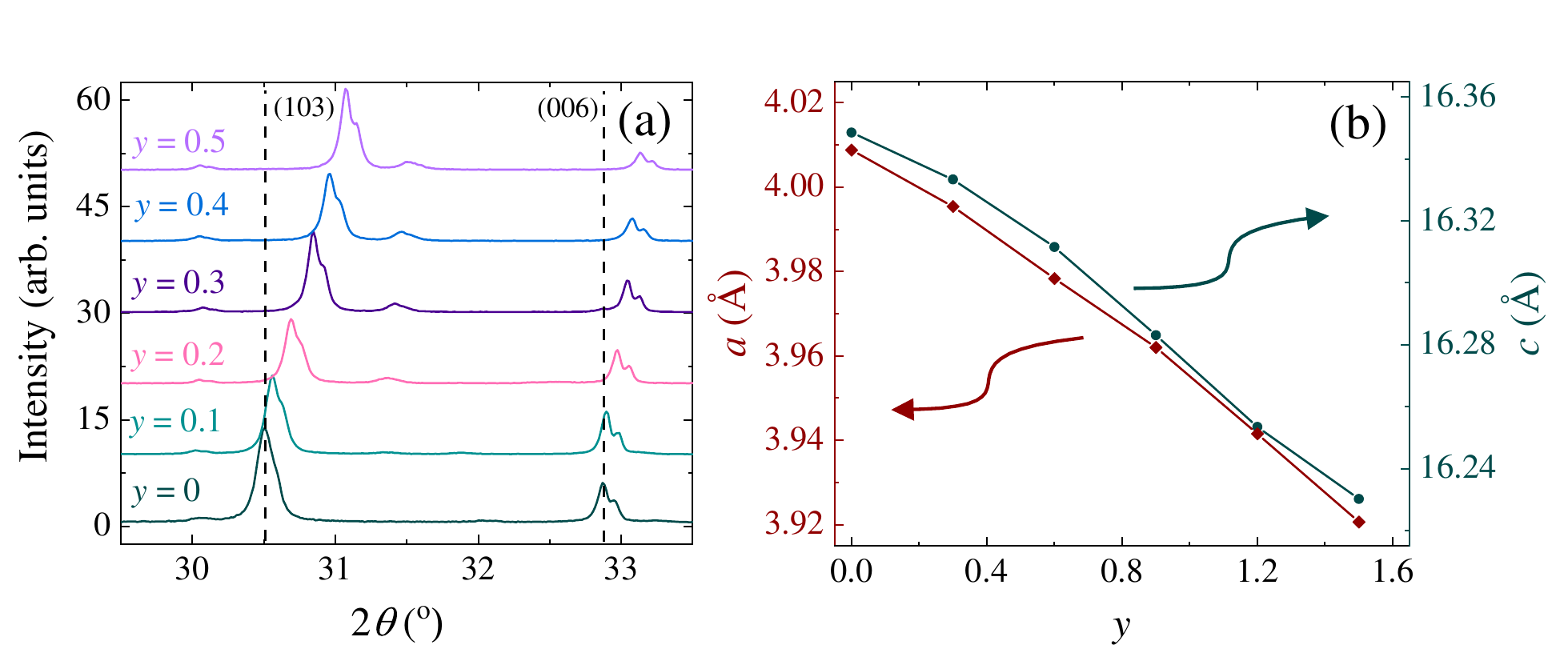}
\caption{(a) Shift of the (103) and (006) diffraction peaks with increasing Co content $y$ in Fe$_{3-y}$Co$_{y}$GeTe$_{2}$. The splitting of the peaks is due to the presence of Cu K$_{\alpha1}$ and K$_{\alpha2}$ radiation in the x-ray source. (b) Variation of the lattice parameters $a$ (left axis, red diamonds) and $c$ (right axis, green circles) with the Co content $y$ in Fe$_{3-y}$Co$_{y}$GeTe$_{2}$.}
\label{FIG: lattice shift}
\end{figure}

The variations in the lattice parameters in the Co substituted powders can be seen to follow a clear trend by tracking the (103) and (006) diffraction peaks as shown in Figure~\ref{FIG: lattice shift}. This gradual shift in the positions of the peaks indicates a decrease in both the $a$ and $c$ lattice parameters, as the level of Co substitution is increased in these materials. 

The dependence of the lattice parameters with Fe composition in Fe$_{3-\delta}$GeTe$_{2}$ has been examined. We show that an Fe deficiency results in a decrease in $a$, while $c$ increases. We find that both the $a$ and $c$ lattice parameters decrease with increasing Co substitution at the Fe site, consistent with what has been reported earlier.~\cite{tian:2019} It was not possible to determine the actual occupancy levels of the Fe/Co in the Co substituted samples using x-ray diffraction, due to the similar atomic numbers and hence x-ray scattering factors of the Fe and Co atoms. Therefore, the decrease observed in the lattice parameters can only be indirectly attributed to the preferential substitution of Co at the 2c (Fe2) sites.

\begin{figure}[t]
\centering
\includegraphics[width=1\textwidth]{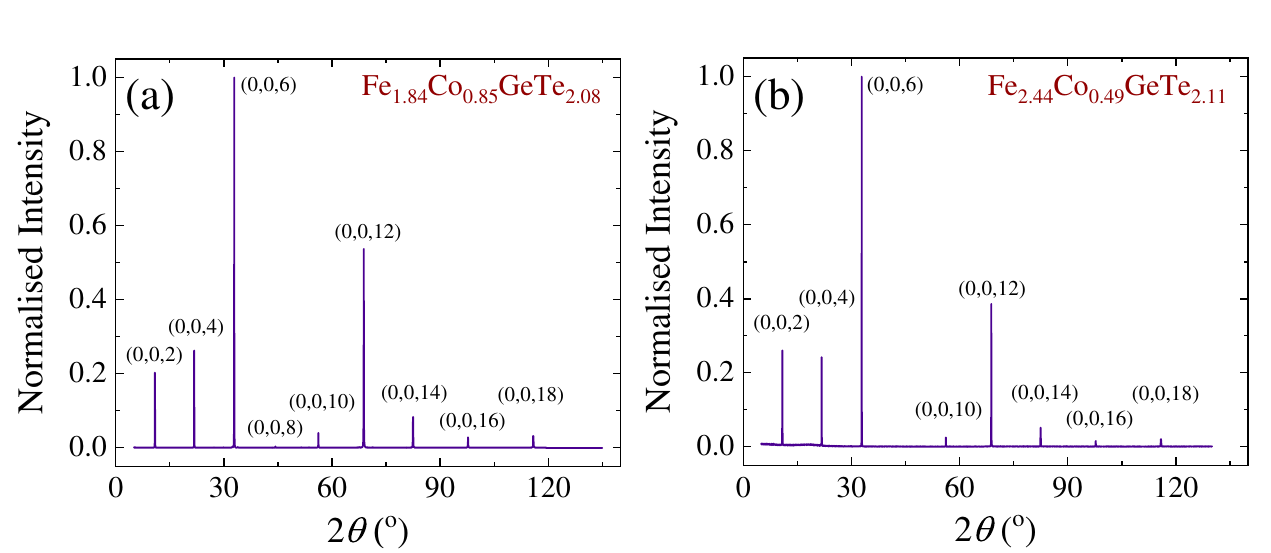}
\caption{X-ray diffraction patterns obtained on single crystals with estimated compositions of (a) Fe$_{1.84}$Co$_{0.85}$GeTe$_{2.08}$ and (b) Fe$_{2.44}$Co$_{0.49}$GeTe$_{2.11}$. The platelets were mounted with the $ab$ plane parallel to the x-ray beam when the scattering angle was zero and the $\left(00l\right)$ reflections observed are shown.}
\label{FIG: xtal XRD}
\end{figure}

X-ray diffraction patterns were also obtained on single crystal platelets mounted with the $ab$ plane parallel to the x-ray beam when the scattering angle is zero, to obtain a series of $\left(00l\right)$ reflections. Figure~\ref{FIG: xtal XRD} shows the diffraction patterns obtained from two different crystal platelets, exhibiting the $\left(00l\right)$ reflections typical of the Fe$_{3}$GeTe$_{2}$ structure. 

\subsection{Energy dispersive x-ray analysis}

\begin{figure}[t]
\centering
\includegraphics[width=\textwidth]{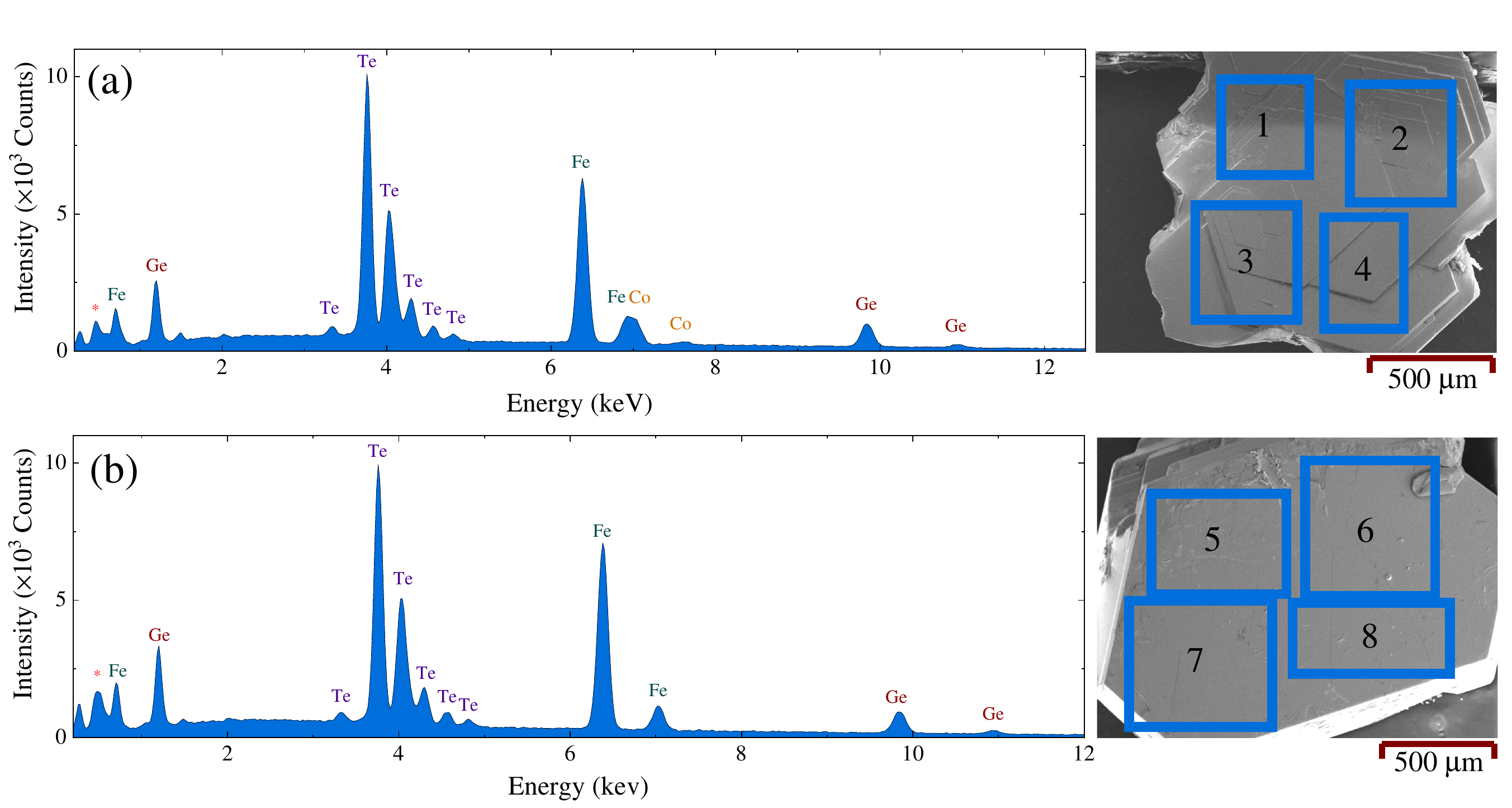}
\caption{EDX spectra for single crystals with nominal starting compositions of (a) Fe$_{2.7}$Co$_{0.3}$GeTe$_{2}$ and (b) Fe$_{3.25}$GeTe$_{2}$ where the size and relative energies of the peaks indicate the elemental composition of the sample. The bulk stoichiometry was measured to be (a) Fe$_{2.44}$Co$_{0.486}$GeTe$_{2.11}$ and (b) Fe$_{2.97}$GeTe$_{2.22}$. Intensity peaks for each element are labeled. Oxygen peaks are indicated by a red asterisk. On the right is a scanning electron microscope image of the surface of each single crystal studied. EDX spectra were collected over the entire area of each crystal. The elemental stoichiometry at each numbered site and across the bulk are given in Table.~\ref{TAB: EDX sites}}
\label{FIG: EDX}
\end{figure}

Composition analysis of the crystals obtained was carried out by EDX analysis. Estimates of the relative Fe, Co, Ge and Te content in the crystals are given in Table~\ref{Tab: EDX}. Examples of the EDX spectra and scanning electron microscope images of two crystals are shown in Figure~\ref{FIG: EDX}. The composition analysis reveals that the Fe content in the crystals is consistently lower than the nominal starting compositions for the various growths.

Crystal growths of unsubstituted Fe$_{3}$GeTe$_{2}$ were attempted while varying the starting Fe content. For crystal growths starting with a nominal Fe content of 3.25 the resulting estimated Fe content in the crystals is close to 3. This indicates that an excess of Fe in the starting materials is necessary to ensure that a maximum occupancy of the Fe sites is achieved. For crystal growths starting with a particular Fe content, it was found that the Fe content in the resulting crystals varied considerably, providing crystals with a range of compositions. An indication of the typical range of final compositions for a nominal starting Fe content of 3.0 is given in the top half of Table~\ref{Tab: EDX}. 
For the Co substituted crystals obtained by both the flux method and CVT growths, the resulting composition of the crystals differed considerably from the nominal starting compositions. Each growth produced a number of crystals which, when examined, showed varying Co content. This spread of Co levels estimated for a few typical crystals obtained from crystal growths with two starting Co substitution levels ($y= 0.3$ and 0.75) are shown in Table~\ref{Tab: EDX}. These resulting compositions are further affirmed by the correlation with the $T_{\mathrm{c}}$ measured on these crystals through magnetic susceptibility measurements as discussed in the next section.

\subsection{Magnetic susceptibility versus temperature} \label{SEC: Magnetisation}

\begin{figure}[t]
\centering
\includegraphics[width=0.5\textwidth]{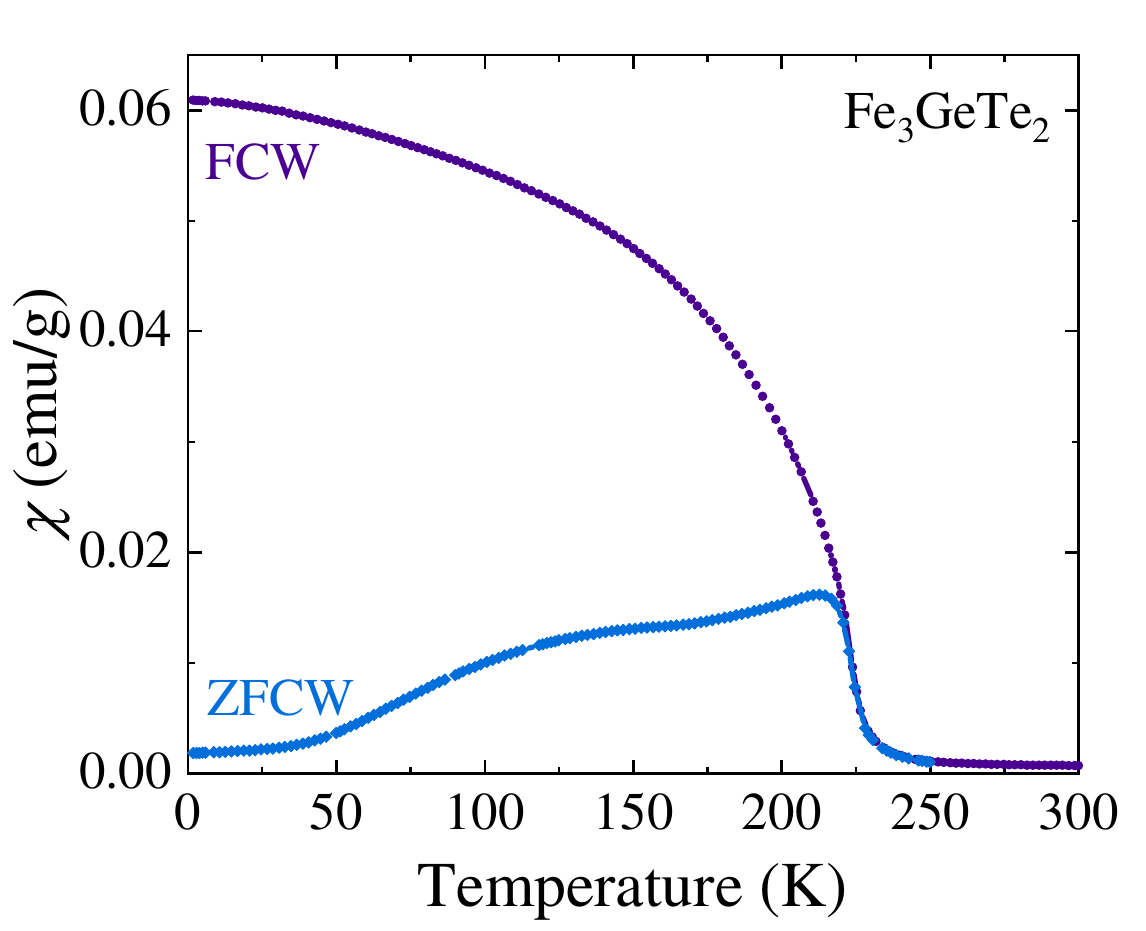}
\caption{Zero-field-cooled warming (ZFCW) and field-cooled-warming (FCW) dc magnetic susceptibility versus temperature for a polycrystalline sample with a nominal composition Fe$_{3}$GeTe$_{2}$ in an applied field of 100~Oe.} 
\label{FIG: Magnetisation Powder}
\end{figure}

The Curie temperature, $T_{\mathrm{c}}$, of a polycrystalline sample with a nominal composition Fe$_{3}$GeTe$_{2}$ was determined from the temperature dependent dc magnetic susceptibility $\chi\left(T\right)$ as shown in Figure~\ref{FIG: Magnetisation Powder} and gave a $T_{\mathrm{c}} = 223(1)$~K. This transition is not as sharp as observed in the single crystals, however, the results are in agreement with previously reported magnetization data for polycrystalline Fe$_{3}$GeTe$_{2}$.~\cite{may:2016}

\begin{figure}[t]
\centering
\includegraphics[width=0.9\textwidth]{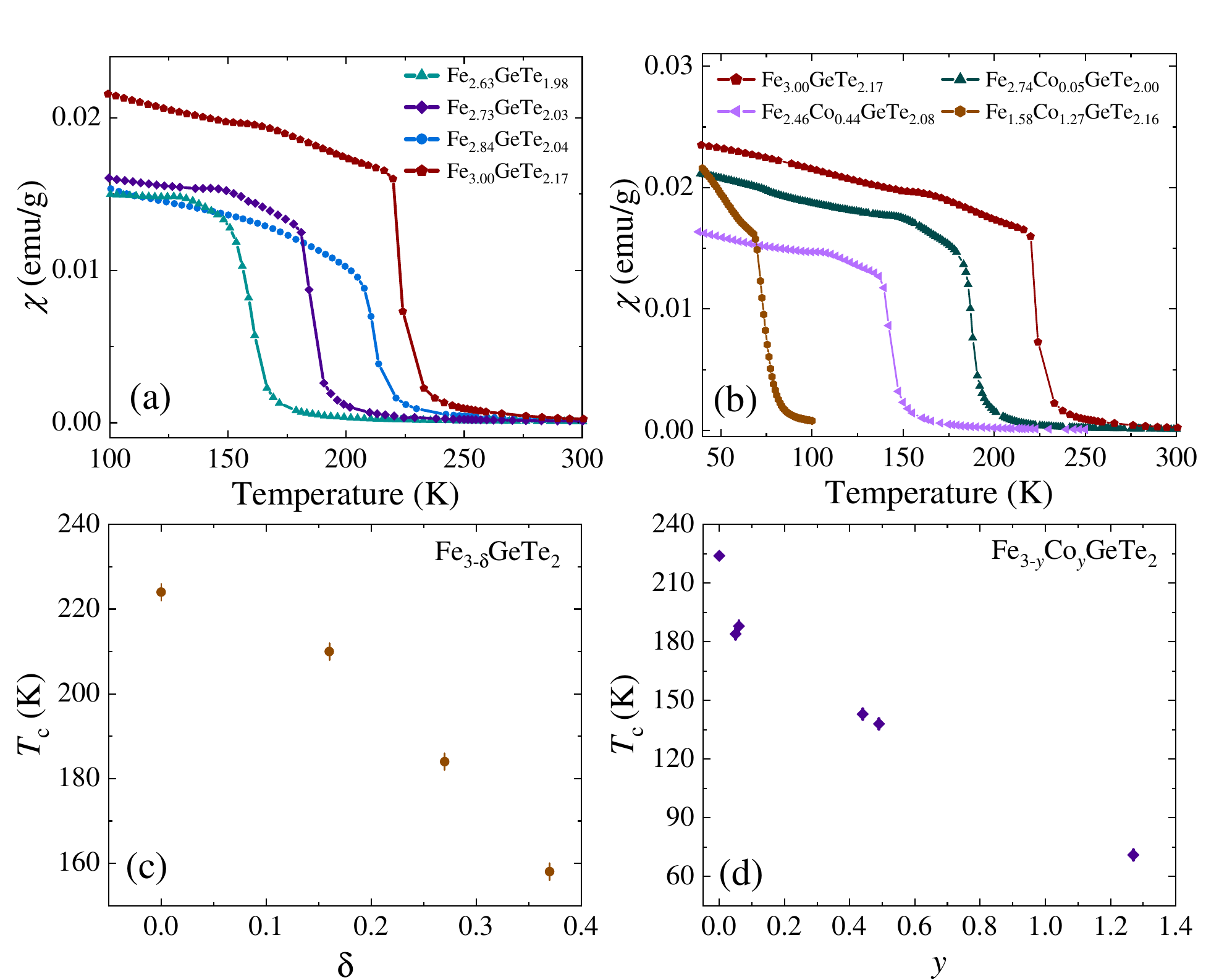}
\caption{(a) Magnetic susceptibility as a function of temperature for several crystals of Fe$_{3-\delta}$GeTe$_{2}$ with differing levels of Fe content. Data were collected  in the field-cooled cooling mode with $H = 100$~Oe applied along the $c$~axis. (b) Magnetic susceptibility as a function of temperature for several crystals of Fe$_{3-y}$Co$_{y}$GeTe$_{2}$ with differing levels of Co content. Data were collected in zero-field-cooled warming mode in a field of 100~Oe applied along the $c$~axis. (c) Transition temperature of Fe$_{3-\delta}$GeTe$_{2}$ as a function of Fe deficiency $\delta$. (d) Transition temperature of Fe$_{3-y}$Co$_{y}$GeTe$_{2}$ as a function of Co content $y$.}
\label{FIG: Magnetisation}
\end{figure}

Field-cooled-cooling dc magnetic susceptibility versus temperature curves for single crystals obtained from CVT with nominal Fe starting compositions of 3 and 3.25 are shown in Figure~\ref{FIG: Magnetisation}. The compositions given in the legend are estimated from EDX. The magnetic susceptibility confirms a ferromagnetic ordering in Fe$_{3-\delta}$GeTe$_{2}$ indicated by the sharp increase in $T_{\mathrm{c}}$ due to spontaneous magnetization at the transition temperature. Crystals from the various crystal growth experiments carried out have a range of Fe content, which is reflected in the range of $T_{\mathrm{c}}$'s they exhibit. Fe deficiency in Fe$_{3-\delta}$GeTe$_{2}$ leads to both a disorder and magnetic dilution, and therefore the highest $T_{\mathrm{c}}$ is obtained when the composition of the resulting crystals is close to Fe$_{3}$GeTe$_{2}$. A rather striking linear dependence of the magnetic transition temperature over a range of Fe content in the Fe$_{3}$GeTe$_{2}$ crystals obtained is shown in Figure~\ref{FIG: Magnetisation}(c).  
The magnetic susceptibility versus temperature behavior in Co substituted Fe-Ge-Te is shown in Figure~\ref{FIG: Magnetisation}(b). Again the compositions given in the legend are those estimated by EDX. The magnetic susceptibility of Fe$_{3}$GeTe$_{2}$ is included for comparison. The transition temperatures of the Co substituted  Fe$_{3}$GeTe$_{2}$ are shown in Figure~\ref{FIG: Magnetisation}(d). Ferromagnetism is preserved for all the Co substitutions attempted here (up to $y =1.27$), with an almost linear decrease in $T_{\mathrm{c}}$ with increasing Co levels..

\section{Summary and conclusions}

The layered magnetic material, Fe$_{3}$GeTe$_{2}$, has been investigated in both polycrystalline and single crystal form. Polycrystalline powders of Fe$_{3}$GeTe$_{2}$ and Fe$_{3-y}$Co$_{y}$GeTe$_{2}$ with several nominal starting compositions and Co substitutions have been synthesized. The effect of varying the nominal starting Fe content on the resulting Fe compositions in the Fe$_{3-\delta}$GeTe$_{2}$ obtained has been studied. We have also produced high quality single crystals with varying Co content ranging from $y = 0$~to~1.5 by both CVT and the flux method. The phase formation, composition analysis and the crystal structures have been investigated using EDX, and both powder and single crystal x-ray diffraction techniques. Our results demonstrate that CVT consistently produces Fe deficient crystals when starting with nominal stoichiometry of Fe$_{3}$GeTe$_{2}$, while starting with an increased Fe content results in crystals with Fe stoichiometry close to the desired Fe$_{3}$. With the crystal growth of the Co substituted crystals by the flux method, starting with a fixed value of $y$ was found to result in crystals exhibiting a range of Co substitutions. The fortuitous range of materials with varying levels of both Fe and Co content produced in this study has allowed us to investigate the dependence of the lattice parameters as well as the magnetic transition temperatures in these systems. The magnetic transition temperatures of all the crystals synthesized have been measured and we find a strong correlation of the transition temperature with the estimated Fe content, with compositions close to the optimum of 3 exhibiting the highest $T_{\mathrm{c}}$. Similarly, the suppression of the magnetic transition temperature with Co substitution has been found. The systematic study of the structural and magnetic properties of these crystals are of great help in understanding the correlations between the synthesis, the structural and magnetic properties in this interesting magnetic material. Such a study is especially important as materials with $T_{\mathrm{c}}$s close to room temperature are seen as being ideal for their applicability to device fabrication. Further studies probing the magnetic properties, especially in relation to their potential to exhibit skyrmions, are currently being explored on these crystals, for which they are ideal. Fe$_{3}$GeTe$_{2}$ continues to remain one of the most interesting 2D magnetic materials with great promise for applications.

\begin{acknowledgement}

We would like to acknowledge Tom Orton, Patrick Ruddy and Daisy Ashworth for their technical support. We would also like to thank David Walker for their assistance with the powder x-ray diffraction and Steve York for their assistance with the energy-dispersive x-ray spectroscopy measurements. This work was financially supported by two Engineering and Physical Sciences Research Council grants: EP/T005963/1, and the UK Skyrmion Project Grant EP/N032128/1. 

\end{acknowledgement}

\bibliography{FGT_bibliography.bib}


\newpage

\begin{table}[tb]
\caption{Nominal starting compositions along with the eventual compositions determined by EDX analysis for single crystals of Fe-Ge-Te and Fe-Co-Ge-Te grown by either CVT or the flux method.} \label{Tab: EDX}
\begin{center}
\begin{tabular}{l l c} 
\hline \hline
Nominal & Composition &  Growth  \\
Composition &  from EDX & Technique  \\
\hline
  Fe$_{2.0}$GeTe$_{4}$ & Fe$_{2.59}$GeTe$_{2.08}$ & Flux \\
  Fe$_{3.0}$GeTe$_{2}$ & Fe$_{2.63}$GeTe$_{1.98}$ & CVT \\
  Fe$_{3.0}$GeTe$_{2}$ & Fe$_{2.73}$GeTe$_{2.03}$ & CVT \\
  Fe$_{3.0}$GeTe$_{2}$ & Fe$_{2.84}$GeTe$_{2.04}$ & CVT \\
  Fe$_{3.0}$GeTe$_{2}$ & Fe$_{2.96}$GeTe$_{2.18}$ & CVT \\
  Fe$_{3.25}$GeTe$_{2}$ & Fe$_{3.00}$GeTe$_{2.17}$ & CVT \\
\hdashline 
  Fe$_{2.00}$Co$_{0.60}$GeTe$_{4}$ & Fe$_{1.88}$Co$_{0.90}$GeTe$_{2.18}$ & Flux \\
  Fe$_{2.00}$Co$_{0.60}$GeTe$_{4}$ & Fe$_{1.84}$Co$_{0.85}$GeTe$_{2.08}$ & Flux \\
  Fe$_{2.10}$Co$_{0.90}$GeTe$_{2}$ & Fe$_{1.60}$Co$_{1.10}$GeTe$_{1.76}$ & CVT \\
  Fe$_{2.10}$Co$_{0.90}$GeTe$_{2}$ & Fe$_{1.65}$Co$_{0.99}$GeTe$_{1.74}$ & CVT \\
  Fe$_{2.25}$Co$_{0.75}$GeTe$_{2}$ & Fe$_{1.58}$Co$_{1.27}$GeTe$_{2.16}$ & CVT \\
  Fe$_{2.25}$Co$_{0.75}$GeTe$_{2}$ & Fe$_{1.67}$Co$_{1.08}$GeTe$_{1.96}$ & CVT \\
  Fe$_{2.25}$Co$_{0.75}$GeTe$_{2}$ & Fe$_{2.74}$Co$_{0.05}$GeTe$_{2.00}$ & CVT \\  
  Fe$_{2.25}$Co$_{0.75}$GeTe$_{2}$ & Fe$_{2.78}$Co$_{0.06}$GeTe$_{1.96}$ & CVT \\
  Fe$_{2.70}$Co$_{0.30}$GeTe$_{2}$ & Fe$_{2.44}$Co$_{0.48}$GeTe$_{2.11}$ & CVT \\
  Fe$_{2.70}$Co$_{0.30}$GeTe$_{2}$ & Fe$_{2.47}$Co$_{0.44}$GeTe$_{2.08}$ & CVT  \\
 \hline \hline
\end{tabular}
\end{center}
\end{table}

\begin{table}[t]
\caption{Crystallographic parameters obtained from a Rietveld refinement of the powder x-ray diffraction pattern collected at room temperature for a polycrystalline sample with a nominal composition Fe$_{3}$GeTe$_{2}$.} \label{Tab: pxrd FGT}
    \centering
        \begin{tabular}{|l| l| l| l| l| l|}
        \hline
        \multicolumn{3}{|l}{Nominal composition} & \multicolumn{3}{|l|}{Fe$_{3}$GeTe$_{2}$} \\
        \multicolumn{3}{|l}{Refined composition} & \multicolumn{3}{|l|}{Fe$_{2.84}$GeTe$_{2}$} \\
        \multicolumn{3}{|l} {Structure} & \multicolumn{3}{|l|}{Hexagonal} \\
        \multicolumn{3}{|l}{Space group} & \multicolumn{3}{|l|}{$P6_{3}/mmc$ (No. 194)} \\
        \multicolumn{3}{|l}{Formula units/unit cell ($Z$)} & \multicolumn{3}{|l|}{2} \\
        \multicolumn{3}{|l}{Lattice parameters} & \multicolumn{3}{|l|}{}\\
        \multicolumn{3}{|l}{$a$ (\AA)} & \multicolumn{3}{|l|}{4.0129(1)} \\
        \multicolumn{3}{|l}{$c$ (\AA)}  & \multicolumn{3}{|l|}{16.3428(4)} \\
        \multicolumn{3}{|l}{$V_{\mathrm{cell}}$ (\AA$^3$)} & \multicolumn{3}{|l|}{227.919(2)} \\
        \multicolumn{3}{|l}{$\rho$ (g/cm$^{3}$)} & \multicolumn{3}{|l|}{7.0876(3)} \\
        \multicolumn{3}{|l}{$R_{\mathrm{wp}}$ (\%)} & \multicolumn{3}{|l|}{10.42} \\
        \multicolumn{3}{|l}{$R_{\mathrm{Bragg}}$ (\%)} & \multicolumn{3}{|l|}{2.401} \\
        \multicolumn{3}{|l}{GOF} & \multicolumn{3}{|l|}{2.77} \\
        \hline
        Atom & Wyckoff position & Occupancy & $x$ & $y$ & $z$  \\
        \hline
        Fe1 & 4e & 1 & 0 & 0 & 0.6703(3)  \\
        Fe2 & 2c & 0.839(8) & 2/3 & 1/3 & 3/4 \\
        Ge1 & 2d & 1 & 1/3 & 2/3 & 3/4 \\
        Te1 & 4f & 1 & 2/3 & 1/3 & 0.58992(11) \\
        \hline
        \end{tabular}       
\end{table}

\begin{table}[h]
\setlength{\tabcolsep}{18pt}
\caption{Composition across selected sites of single crystals with the nominal starting compositions Fe$_{2.7}$Co$_{0.3}$GeTe$_{2}$ and Fe$_{3.25}$GeTe$_{2}$ as indicated in Figure~\ref{FIG: EDX} determined from energy dispersive x-ray analysis.} \label{TAB: EDX sites}
\begin{center}

\begin{tabular}{lllll} 

\hline
\hline
Spectrum Label & Fe (\%) & Co (\%) & Ge (\%) & Te (\%) \\
 \hline
Fe$_{2.7}$Co$_{0.3}$GeTe$_{2}$ &  &  &  &  \\
\hdashline
1 & 41.05(1) & 7.76(1) & 16.61(1) & 34.58(1) \\
2 & 40.27(1) & 8.16(1) & 16.59(1) & 34.99(1) \\
3 & 40.56(1) & 7.93(1) & 16.72(1) & 34.79(1) \\
4 & 40.10(1) & 8.31(1) & 16.22(1) & 35.37(1) \\
\hdashline
Fe$_{3.25}$GeTe$_{2}$ &  &  &  &  \\
\hdashline
5 & 47.90(1) & 0 & 16.16(1) & 35.94(1) \\
6 & 47.44(1) & 0 & 16.39(1) & 36.17(1) \\
7 & 48.35(1) & 0 & 15.61(1) & 36.04(1) \\
8 & 48.13(1) & 0 & 16.38(1) & 35.49(1) \\
\hline

\end{tabular}
\end{center}
\end{table}

\end{document}